\documentstyle[preprint,psfig,aps,prl]{revtex}

\newcommand{\na}{$\rm NaV_2O_5$ }  
\newcommand{\nae}{$\rm NaV_2O_5$}  
\newcommand{\dxy}{$d_{xy}$ }  
\newcommand{\dxys}{$d_{xy}^*$ }  
\newcommand{\be}{\begin{equation}}
\newcommand{\ee}{\end{equation}}
\newcommand{\la}{\langle}
\newcommand{\ra}{\rangle}
\begin{document}

\draft

\title{Electron Correlation Effects in  \\
Resonant Inelastic X-ray
Scattering of NaV$_2$O$_5$}

\author{G. P. Zhang$^{1,2}$, T. A. Callcott$^1$, G. T. Woods$^1$, and
L. Lin$^1$ } 

\address{$^1$Department of Physics and Astronomy, The University of
Tennessee, Knoxville, TN 37996-1200}

\address{$^2$State University of New York, College at Buffalo, Buffalo,
NY 14222}

\author{Brian Sales, D. Mandrus, and J. He}

\address{Solid State Division, Oak Ridge National Laboratory, TN
37831}

\date{\today}
\maketitle

\begin{abstract}

Element- and site-specific resonant inelastic x-ray scattering
spectroscopy (RIXS) is employed to investigate electron correlation
effects in {$\rm NaV_2O_5$}.  In contrast to single photon techniques,
RIXS at the vanadium $L_3$ edge is able to probe $d-d^*$ transitions
between V $d$-bands.  A sharp energy loss feature is observed at $-1.56$
eV, which is well reproduced by a model calculation including
correlation effects.  The calculation identifies the loss feature as
excitation between the lower and upper Hubbard bands and permits an
accurate determination of the Hubbard interaction term $U= 3.0 \pm 0.2 $
eV.

\end{abstract}


Orthorhombic \nae, with symmetry $ P_{mmn}$, is a new type of Mott
insulator and possesses many exotic properties\cite{smo,mar}. It
exhibits a spin-Peirels-like phase transition\cite{japan}, but does
not have some typical features shown by CuGeO$_3$ \cite{bom} such as
change of the transition temperature with magnetic fields
\cite{vasil}.  As in many Mott insulators, the strong correlation in
\na splits its original $d_{xy}$ band into two new Hubbard subbands
$d_{xy}$ and $d_{xy}^*$, and opens a gap around the Fermi
surface. Such a gap is strong evidence of the electron correlation
effect.  It probably cannot be probed optically; due to the dipole
selection rule, a direct transition between \dxy and \dxys is very
weak if not completely forbidden. In particular, recent work pointed
out that the 1.0 eV peak in the optical absorption spectrum does not
originate from the so-called $d-d$ transition as previously
suggested\cite{go}, but is related to the number of V$^{5+}$
ions\cite{mar}.  The correlated nature of \na was also investigated by
angular-resolved-photoemission spectroscopy (ARPES)\cite{kob}, but the
results may be compromised by charging effects.  Even assuming no
charging effect, ARPES probes only the occupied states with little
information on the unoccupied bands\cite{hasan}.  Electron energy-loss
spectroscopy (EELS) faces a similar difficulty as ARPES imaging only
the unoccupied states.  Furthermore, the elastic contribution and
surface losses make EELS unable to measure at zero momentum
transfer\cite{hasan,fink}.  To this end, a direct probe of the
correlation effect is yet to come.

Element- and site-specific resonant inelastic x-ray scattering
spectroscopy (RIXS) \cite{tom1} is a powerful tool to
{\it directly} probe those optically {\it inaccessible} bands in
complex materials. It can project out the density of states element by
element, which can be directly compared to the theoretical
calculation.  In this Letter, we use RIXS to directly probe electronic
excitations across the Fermi level in \nae.  The RIXS spectra taken at
the vanadium $L_3$ edge show two principal features. A broad band
located between $-4$ and $-10$ eV on the energy loss scale derives
from strongly hybridized V-$d$/O-$p$ bands.  A smaller energy loss
feature at $-1.56$ eV, which resonates strongly in a narrow range of
incident photon energies near threshold, is the focus of this paper.
It is associated with a narrow band derived predominantly from
V-$d_{xy}$ orbitals with little contribution from O-$p$ orbitals. In
the absence of correlation effects, the band spans the Fermi level.
Correlation effects split the band into upper and lower Hubbard bands.
Because of the lack of O-$p$ hybridization with the vanadium orbitals
creating this band, we are able to use a simplified model that
suppresses the V-O interactions but accurately describes the \dxy band
at the Fermi level and accounts for correlation effects between
vanadium atoms. The model calculation accurately reproduces the
$-$1.56 eV energy loss feature with a Hubbard interaction term of $U =
3.0\pm 0.2$ eV, and identifies this energy loss with a $d-d^*$
excitation between the lower and upper Hubbard bands.

Resonant inelastic x-ray scattering is a photon-in and photon-out
technique shown schematically in Fig. 1.  An incoming photon first
excites electrons from the core level to the unoccupied bands.  Since
the core binding energy is unique to one particular element, by using
different photon energies one can selectively excite a desired
element. Thus, this is an element-specific technique.  In the decay
process, an electron from occupied bands recombines with the core
hole, leaving an electron in the unoccupied bands and a hole in the
occupied bands. The energy loss between the incident photon and
emitted photon energies measures the difference between the occupied
and unoccupied bands. The smallest energy-loss peak corresponds to the
lowest excitation between states with the angular momentum change of
$\Delta l=0,\pm 2$. If the excited electron directly recombines with
the core hole without involving the valence electrons, this gives an
elastic peak whose energy is exactly equal to the incident photon
energy.

Single crystals of $\rm{NaV_{2}O_{5}}$ were prepared at the Oak Ridge
National Lab.  The samples were characterized by x-ray diffraction,
specific heat and magnetic susceptibility measurements, and were
cleaved along the $c$ direction to provide clean and flat [001]
surfaces containing the $a$ and $b$ axes. The long dimension was along
the $b$ axis and the shorter dimension along the $a$ axis.  Our
measurements were performed at the soft x-ray spectroscopy endstation
on undulator Beam-line 8.0 at the Advanced Light Source (ALS) located
at the Lawrence Berkeley National Lab. Our monochromator was
calibrated with Ti-$L$ edge of TiO$_2$ and compared with the V-$L$
edge of $\rm V_2O_5$; the spectrometer was calibrated according to the
elastic peak.  Monochromatized light from the undulator is incident on
samples placed within a few mm of the entrance slit of an emission
spectrometer.  The incident beam is $p$-polarized and the emission
spectrum is taken at 90$^{\circ}$ from the incident beam.

Figure 2(a) shows the V-$L_{3}$ emission spectra versus emitted photon
energies. The incident photon energies are labeled from a to g.  The
vertical bars mark the position of the elastic peak located at the
incident photon energy. The excitation energies are also indicated on a
plot of the total fluorescent yield in Fig. 2(b).  The emission spectra
display two main features: peak A on the left-hand side and peak B on
the right-hand side.  At energies below about 516 eV, both peaks track
the incident photon energy to higher energies in the way characteristic
of RIXS spectra, where there is a fixed energy separation between
incident and emitted photons\cite{tom4}.  Above about 516 eV, the
spectra evolve into soft x-ray normal fluorescence (SXF) spectra in
which incident and emitted photons are decoupled and the spectral
features are fixed in energy.  In the transition region, both RIXS and
SXF features may be present in the spectra as is observed for peak B for
incident energies of 518 and 519 eV.  Peak A results from the hybridized
oxygen $2p$ and vanadium $3d$ bands (see the lower left part of Fig. 3).
Elsewhere\cite{gw}, we explore in more detail the density of states
information from absorption and SXF spectra taken at both the V-$L_3$
and the O-$K$ edges.  Here we concentrate on the particular features
associated with peak B in the vicinity of the threshold at 516 eV, which
provide specific information on correlation effects of the vanadium
atoms in this material.

In Fig. 2(c), an enlarged view of peak B is shown on an energy loss
scale which is calculated by subtracting the elastic peak energies from
the inelastic energies on the same emission spectra.  In the RIXS region
below 516 eV (a-d), a fixed energy loss of $-(1.56 \pm 0.05)$ eV is
observed.  The peak resonates strongly at 516 eV (d).  As may be seen
from Eq. 2 below, the resonance occurs when the energy of incident
photons is just sufficient to excite a 2$p$ core electron into a real
intermediate state.  In the present case we believe that this resonance
excitation occurs to V-$d_{xy}$ states derived from the upper Hubbard
band, with emission occurring from correlated lower Hubbard band states.
It is important to note that, although the position and strength of the
resonance is effected by core-hole localization in the intermediate
state, the observed energy losses involve only the initial and final
states of the scattering process.  Thus, the measured energy losses do
not include the effects of core hole localization and lifetime, and are
an accurate measure of the $d-d^*$ excitation between the lower and
upper Hubbard bands. In Fig. 3, we illustrate our picture for
interpreting the spectra.  Projected to the left is the total density of
states calculated using the {\sc Wien97} band structure
codes\cite{wien}.  The lower valence band is formed primarily from
V-$3d$ states hybridized with O-$2p$ states. The band spanning the Fermi
level is formed from nearly pure V-$d_{xy}$ orbitals that interact along
the rungs of the ladder-like structure of \nae.  The bands above the
Fermi level are also formed from strongly hybridized V-$d$/O-$p$
orbitals.  Projected to the right in the figure is our picture of the
spectra when correlation is added.  The V-$d_{xy}$ band spanning the
Fermi level is now split by correlation effects, with the lowest
observed energy loss corresponding to the splitting of this band.

Such correlation effects can be better understood within a simple
model cluster. In order to do so, we choose a ladder consisting of
eight V atoms with a periodic boundary condition along $b$ axis as
shown in Fig. 4. Only V-$d_{xy}$ orbitals around the Fermi surface are
explicitly taken into account.  Oxygen orbitals are suppressed though
their presence is partially taken into account by fitting the results
to a band calculation.  This is a good approximation for the present
problem since the V-$d_{xy}$ band dominates around the Fermi surface
(see Fig. 3) and there is little contribution from the oxygen
bands\cite{gw}. Therefore, we focus on peak B in the experimental
curve (see Fig. 2). 

Within these approximations, we write the Hamiltonian
as\cite{fink,note1,horsch,sa} \be
H=-\sum_{<ij>\sigma}t_{ij}^{a(b)}(c^{\dagger}_{i\sigma}c_{j\sigma}+h.c.)+U\sum_in_{i\uparrow}n_{i\downarrow}+\sum_{\mu\in
a,b}V^{\mu}\sum_in_i^{\mu}n_{i+1}^{\mu}~, \ee where the nearest neighbor
hopping integral $t^{a(b)}_{<ij>}$ along the $a(b)$ direction is
obtained by fitting the {\it ab initio} band structure\cite{smo}; the
on-site interaction $U$ is from our experiment; the intersite electron
correlation along the $a(b)$ direction, $V^{a(b)}$, is obtained from a
previous optical study\cite{fink,horsch,sa}. To be more specific, we
choose $t^a=0.38$, $t^b=0.17$, $U=3$, $V^a=0.72$ and $V^b=0.72$ eV. The
number of electrons of the system is $N=4$.  All of the operators are
standard\cite{horsch}.  Within such a small cluster, we can directly
diagonalize the many-body Hamiltonian with the Lanczos recursion method
\cite{zhang} and obtain eigenstates and eigenvalues. The RIXS spectrum
is calculated from the Kramers-Heisenberg formula \cite{tom1,tom4} \be
S(\omega,\omega')=\sum_f\left | \sum_m\frac{\la f| p\cdot A |m\ra\la
m|p\cdot A|gs\ra}{\omega+E_{gs}-E_m-i\Gamma}\right |^2
\delta(E_{gs}+\omega-E_f-\omega')~, \ee where $|gs\ra$, $|m\ra$, and
$|f\ra$ are initial, intermediate, and final states, and $E_{gs}$,
$E_m$, and $E_f$ are their energies, respectively; $\omega$ and
$\omega'$ are the incident and emitted photon energies; $\Gamma$ is the
spectral broadening due to the core lifetime in the intermediate
state. $p\cdot A$ is the transition operator, which is approximated by a
dipole operator. Our approach is similar to a previous
calculation\cite{kotani1}.  In the intermediate states, the system
contains $N+1$ electrons and one hole in the core level; in the final
states, the system has $N$ electrons and the core hole is refilled
leaving an electron-hole excitation behind.

In metals, since there is a finite density of states at the Fermi level,
RIXS starts from a zero-energy loss. In \na the LDA calculation catches
almost all the experimental features but predicts a metallic phase
around the Fermi level. Consequently, the LDA-RIXS spectrum begins from
a zero-energy loss, in contrast to our experimental
observation. However, we can recover this missing feature by including
electron correlations.  The results are shown in Fig. 4, where the
intensities are on the same but arbitrary scale, thus they are
comparable.  The incident energy is labeled from a to d with a strong
resonance in curve c. In the inset, incident energies are also marked,
where the core level energy is already subtracted from the energy scale.
Upon increase in the incident photon energy, the spectrum exhibits a
familiar inelastic resonance feature as we have seen in our experiments:
The peak tracks the incident photon energy up to the threshold.  The
prominent feature is a resonant peak which appears at an energy loss of
$-1.55$ eV (see the arrow), which can be directly compared with our
experimental value of $-1.56$ eV.  Varying $U$ in the calculation
produces a linear change in this energy loss feature.  We find that a
variation in $U$ of $3.0\pm 0.2$ eV corresponds to the experimental
uncertainty of the energy loss of $-(1.56 \pm 0.05)$ eV.  Previous
theoretical and experimental studies have estimated $U$ ranging from 2
to 7 eV\cite{smo,mar,fink,horsch,sa,newref}.  We believe that our
value is soundly based on an appropriate model accurately describing the
process being precisely measured experimentally, and thus strongly
constrains $U$ to values near 3.0 eV.

Two other weak resonances are apparent in the theoretical emission
spectra at about -2.8 eV and -1.2 eV, respectively.  The former one may
indicate the continuum edge, but in order to determine this edge
accurately, a much larger system which accurately includes V-O
interaction is needed. The latter one mainly results from the finite
size effect of the cluster used in the calculation. We believe that it
becomes relatively weaker as the cluster size is increased and becomes
negligible for large clusters.  As aforementioned, since the lowest
energy loss of the resonant inelastic peak measures the lowest
excitation energy, the nonzero energy loss unambiguously establishes a
gap is opened. However, if we switch off the electron correlation terms,
from Eqn. 1 we know that this comes back to the LDA results where the
gap is closed and consequently the resonant peak starts from the
zero-energy loss. This shows the importance of the electron correlation
effect. Since around the Fermi level, the density of states is
predominantly of \dxy characters, the gap must be between the lower
Hubbard \dxy band and the upper Hubbard $d_{xy}^*$ band.  Therefore, our
experiment and theory jointly point out an old wisdom\cite{smo,hubbard}
in \nae: the electron correlation splits the original V-$d_{xy}$ bands
and drives the system into an insulating phase.

In addition, a decade of experimental work \cite{tom4} has shown that
the RIXS spectra have intensities of the similar magnitude as ordinary
SXF spectra since theoretically the matrix elements governing RIXS and
SXF are substantially the same except for modifications of spectral
shape due to interference effects and enhancements in RIXS spectra due
to the resonant denominator.  Using the ratio of the RIXS intensity to
the SXF intensity, we can estimate the scattering cross-section of RIXS.
We found that depending on incident photon frequencies, the ratio is
about 2--3 times in our experiment, which is compatible with our
theoretical estimation of 2--4 times.

In conclusion, our element- and site-specific resonant inelastic x-ray
scattering spectrum provides much insight into the complicated
electron structure in \nae. Electron correlation manifests itself by
opening the gap around the Fermi surface.  Our study demonstrates the
value of soft x-ray spectroscopies for directly probing the
correlation effects in these interesting transition-metal oxides
\cite{lee}, high $T_c$ superconductors \cite{hasan} and GMR
materials.

This research is supported by NSF grant DMR-9801804. Samples were
prepared and characterized at ORNL supported by DOE contract
DE-AC05-00OR-22725. Measurements were carried out at the Advanced
Light Source at LBNL supported by DOE contract DE-A003-76SF00098. We
would like to acknowledge Dr. C. Halloy and the staff at the Joint
Institute of Computational Sciences at UT where part of our
computation has been done. Finally, we would like to thank the
referees for their very important and helpful suggestions.

\newcommand{\et}{{\it et al., }}

\begin{figure}
\caption{Mechanism of resonant inelastic x-ray scattering. Note that
the momentum transfer is very small.}
\end{figure}

\begin{figure}
\caption{(a) V$-L_3$ edge resonant inelastic x-ray scattering spectrum
in \nae.  Incident energies are labeled from a to g.  The vertical
bars denote the elastic peak positions. (b) Total fluorescence yield
(TFY).  (c) Enlarged view of peak B on the energy loss scale.  }
\end{figure}

\begin{figure}
\caption{Correlated picture of \nae. Left panel: LDA calculation;
Right panel: Modification due to electron correlation.  }
\end{figure}

\begin{figure}
\caption{Correlated model calculation of RIXS. Intensities are in the
same but arbitrary units. The arrow denotes the
resonant inelastic peak position.  Inset: (left) absorption spectrum;
(right) used model cluster.  }
\end{figure}








\newpage

~~~~~~~~~~~~~~~~
\vspace{1cm}

\begin{center}

\hspace{0.cm}\psfig{figure=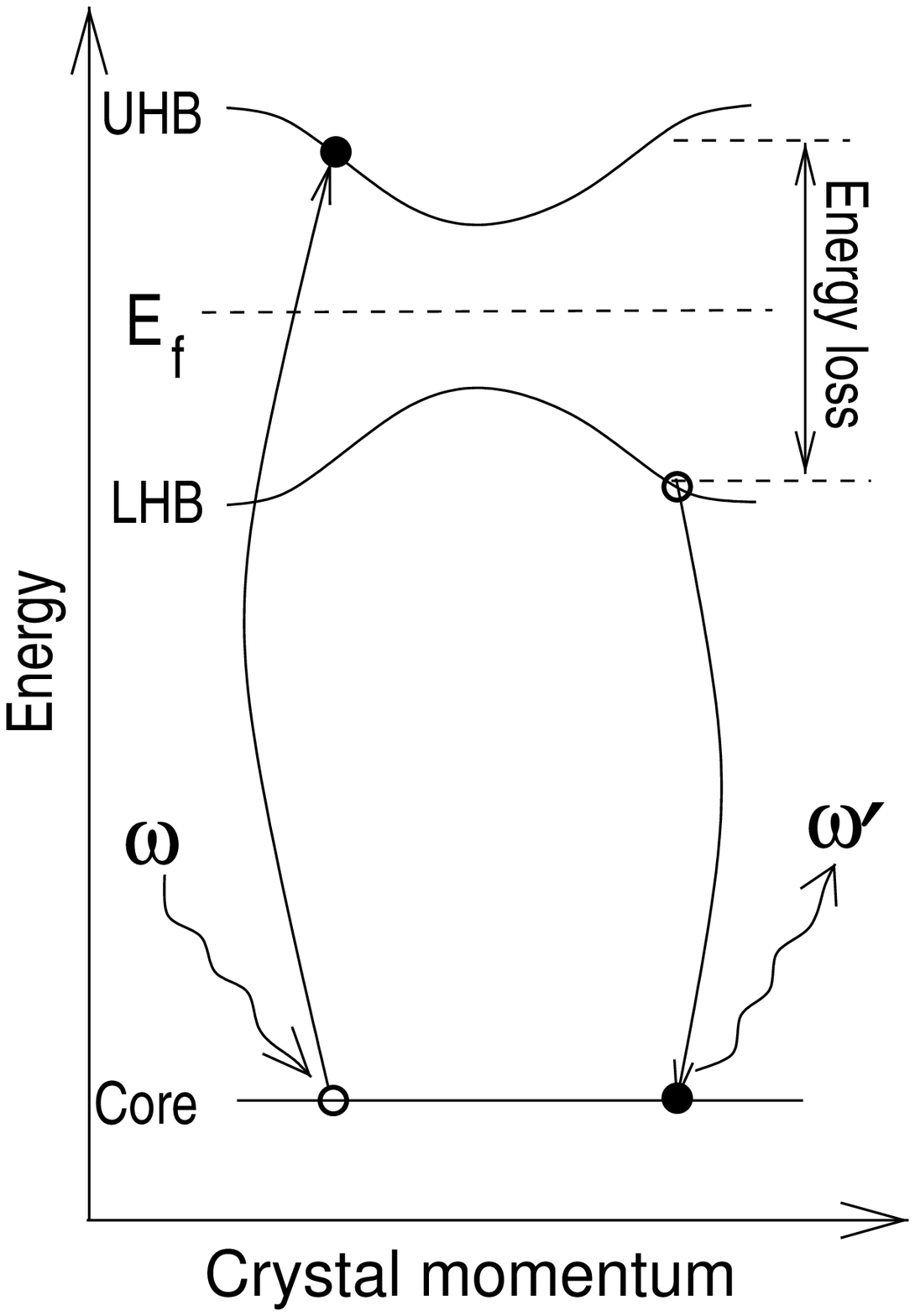,width=10cm,angle=0}

\vspace{3cm}

\centerline{Figure 1}

\newpage

~~~~~~~~~~~~~~~~

\vspace{1cm}

\hspace{-0.8cm}\psfig{figure=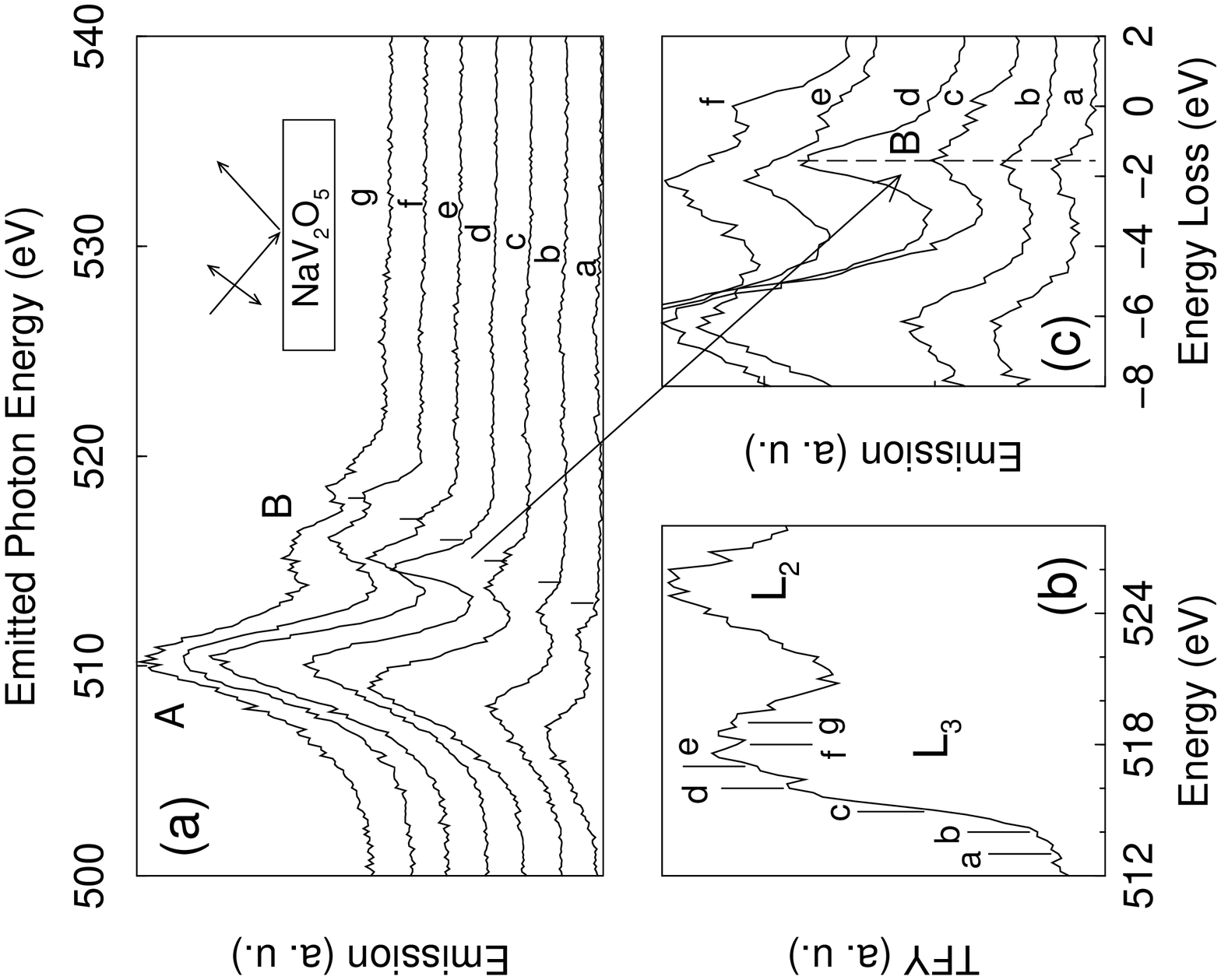,width=12cm,angle=270}

\vspace{3cm}

\centerline{Figure 2}

\newpage

~~~~~~~~~~~~~~~~

\vspace{1cm}

\hspace{0.5cm}\psfig{figure=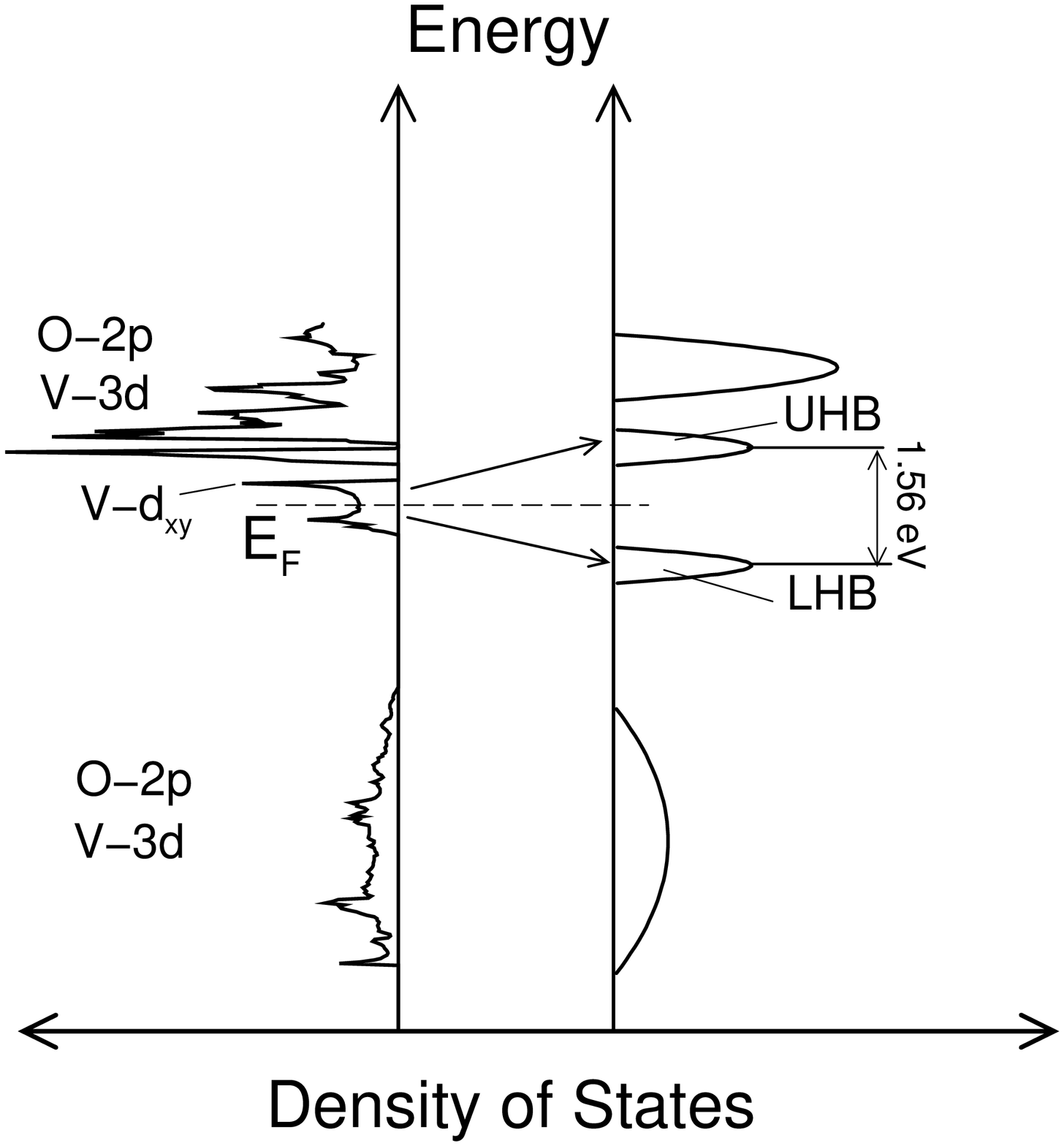,width=11cm,angle=270}

\vspace{4cm}

\centerline{Figure 3}

\newpage
~~~~~~~~~~~~~~~~

\vspace{1cm}

\hspace{-1cm}\psfig{figure=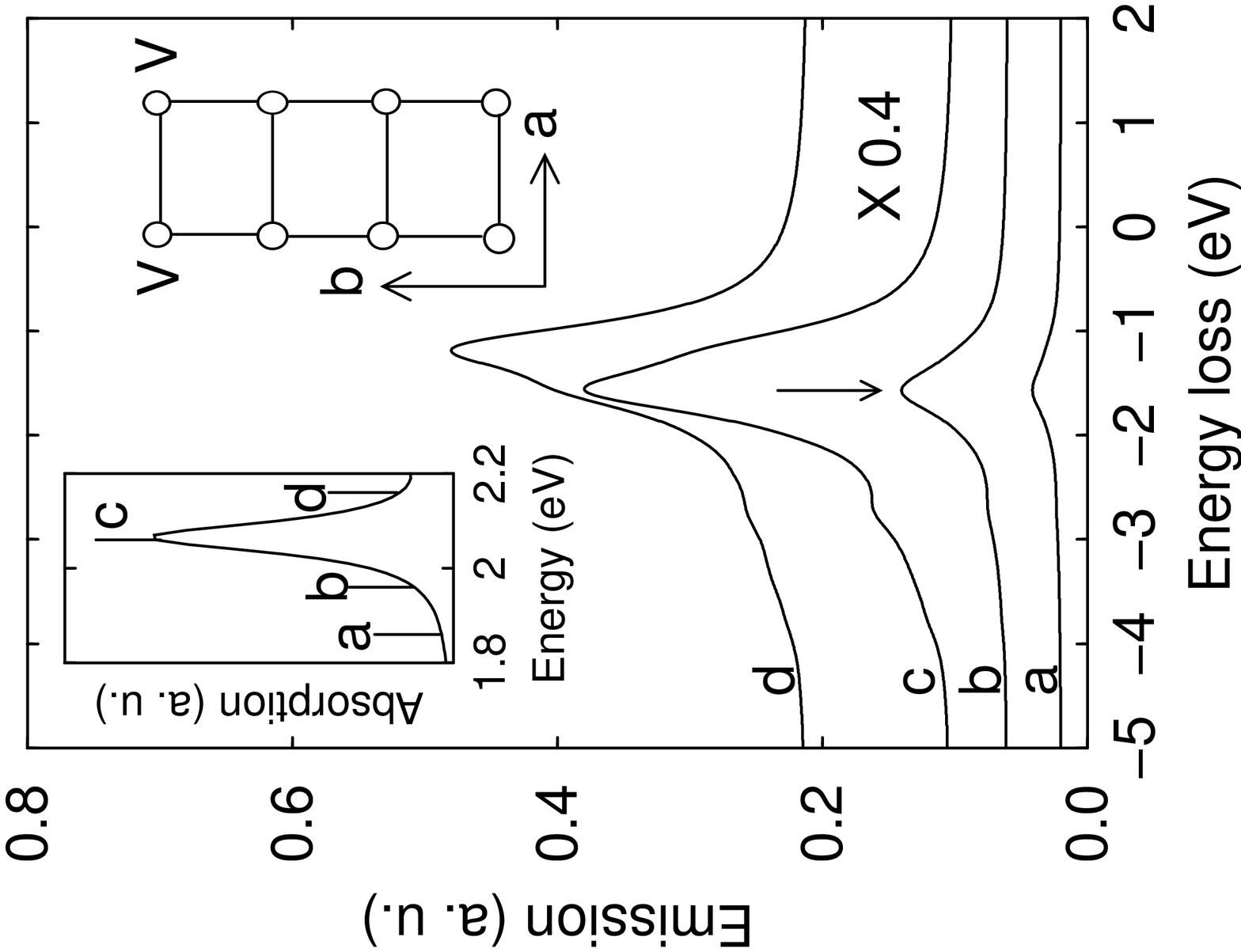,width=11cm,angle=270}

\vspace{4cm}

\centerline{Figure 4}

\end{center}


\begin{references}
\bibitem{note1} The core level term is subtracted from the
Hamiltonian.  The strength of the interaction between the 3$d$
electron and the core hole is unknown at present. Instead of
introducing an unknown quantity into our calculation, we leave this
for a future study.
\end{references}
\end{document}